\documentclass[12pt]{article}

\newcommand{\newsection}{\setcounter{equation}{0}\section}

\def\appendix#1{\addtocounter{section}{1}\setcounter{equation}{0}
\renewcommand{\thesection}{\Alph{section}}
\section*{Appendix\thesection\protect\indent \parbox[t]{11.715cm} {#1}}
\addcontentsline{toc}{section}{Appendix \thesection\ \ \ #1} }
\def\slash{{\!\!\!/\,}} 

\def\nn{\nonumber}

\def\e{{\,\rm e}\,}

\newcommand{\non}{\nonumber \\}

\def\be{\begin{equation}}
\def\ee{\end{equation}}
\def\bea{\begin{eqnarray}}
\def\eea{\end{eqnarray}}
\def\bd{\begin{displaymath}}
\def\ed{\end{displaymath}}

\makeatletter
\newdimen\normalarrayskip              
\newdimen\minarrayskip                 
\normalarrayskip\baselineskip
\minarrayskip\jot
\newif\ifold             \oldtrue            \def\new{\oldfalse}
\def\arraymode{\ifold\relax\else\displaystyle\fi} 
\def\@arrayskip{\ifold\baselineskip\z@\lineskip\z@
     \else
     \baselineskip\minarrayskip\lineskip2\minarrayskip\fi}
\def\@arrayclassz{\ifcase \@lastchclass \@acolampacol \or
\@ampacol \or \or \or \@addamp \or
   \@acolampacol \or \@firstampfalse \@acol \fi
\edef\@preamble{\@preamble
  \ifcase \@chnum
     \hfil$\relax\arraymode\@sharp$\hfil
     \or $\relax\arraymode\@sharp$\hfil
     \or \hfil$\relax\arraymode\@sharp$\fi}}
\def\@array[#1]#2{\setbox\@arstrutbox=\hbox{\vrule
     height\arraystretch \ht\strutbox
     depth\arraystretch \dp\strutbox
     width\z@}\@mkpream{#2}\edef\@preamble{\halign \noexpand\@halignto
\bgroup \tabskip\z@ \@arstrut \@preamble \tabskip\z@ \cr}%
\let\@startpbox\@@startpbox \let\@endpbox\@@endpbox
  \if #1t\vtop \else \if#1b\vbox \else \vcenter \fi\fi
  \bgroup \let\par\relax
  \let\@sharp##\let\protect\relax
  \@arrayskip\@preamble}
\makeatother

\newcommand{\beq}{\begin{eqnarray}}
\newcommand{\eeq}{\end{eqnarray}}

\newcommand{\G}{\Gamma}

\begin{document}
\begin{titlepage}
\begin{flushright}

\baselineskip=12pt
HWM--01--23\\ EMPG--01--09\\ hep--th/0106259\\
\hfill{ }\\ Revised Version\\ September 2001
\end{flushright}

\begin{center}

\baselineskip=24pt

{\Large\bf D-Brane Dynamics\\ and Logarithmic Superconformal Algebras}

\baselineskip=14pt

\vspace{1cm}

{\bf Nick E. Mavromatos}
\\[3mm]
{\it Department of Physics\\ King's College London\\ Strand, London WC2R 2LS,
U.K.}\\ {\tt Nikolaos.Mavromatos@cern.ch}
\\[6mm]

{\bf Richard J. Szabo}
\\[3mm]
{\it Department of Mathematics\\ Heriot-Watt University\\ Riccarton,
Edinburgh EH14 4AS, U.K.}\\  {\tt R.J.Szabo@ma.hw.ac.uk}
\\[15mm]

\end{center}

\vskip 2 cm

\begin{abstract}

\baselineskip=12pt

We construct the consistent supersymmetric extensions of the operators
describing the recoil of a D-brane and show that they realize an ${\cal N}=1$
logarithmic superconformal algebra. The corresponding supersymmetric vertex
operator is related to the action of a twisted superparticle with twist field
determined by the angular momentum of the recoiling D-brane and with explicitly
broken $\kappa$-symmetry. We show that the superconformal completion removes
the logarithmic modular divergences that are present in the bosonic string loop
scattering amplitudes. These features are all consequences of the relationship
that exists in these models between worldsheet rescaling and the time evolution
of the D-brane in target space.

\end{abstract}

\end{titlepage}

\newsection{Introduction}

One of the ways of describing the dynamics of D-branes is
to regard them as string solitons. The center of mass zero modes of the soliton
break target space translational symmetry. This effect can be described in an
open string worldsheet formalism as inducing a deformation of the underlying
conformal field theory $\sigma$-model~\cite{FPR1}. The change of state of the
soliton during a scattering event leads to a relevant deformation of the
$\sigma$-model and causes a change in the conformal field theory background.
One surprising aspect of this description is that the quantization of the
collective coordinates of the D-brane induces violations of conformal
invariance in the form of non-local worldsheet operators, yielding unexpected
logarithmic terms~\cite{FPR1}. These scaling violations were subsequently
shown~\cite{KM} to be a consequence of the fact that the worldsheet Noether
currents associated with spatial translations should be identified with
logarithmic operators~\cite{Gurarie} of the theory. These logarithmic operators
correspond to hidden (target space) continuous symmetries~\cite{CKT} related to
the collective coordinates. The proper operators describing the recoil effects
during the scattering of closed string states off the D-brane background have
been constructed and their logarithmic conformal algebra is
well-studied~\cite{recoil}--\cite{MS1}.

Logarithmic conformal field theories lie on the border between conformal field
theories and generic renormalizable two-dimensional field theories. They are
characterized by the appearence of logarithmic terms in the four-point
correlation functions of primary operators. The logarithmic divergences are due
to the fact that two (or more) primary fields have the same scaling dimensions
(modulo integers) and that they can no longer diagonalize the Virasoro
generators of the theory~\cite{Gurarie}. Rather, they mix under conformal
transformations and form Jordan cells. While such theories are not unitary,
they define realistic models which can still be classified to some extent by
conformal data. In certain cases, the logarithmic scaling violations lead to
subtleties in their worldsheet renormalization group properties, implying a
sort of marginal non-criticality of the theory. This is precisely the case with
the recoil deformation operators for string solitons, which are marginally
relevant, and are thereby capable of driving the deformed $\sigma$-model to a
non-trivial
fixed point. These operators are described in a semi-classical impulse
approximation, appropriate for the non-relativistic dynamics of heavy D-branes,
or equivalently to leading tree-level order of string perturbation theory.
It is induced by the capture of a closed string state by the D-brane defect. In
the worldsheet formalism such a procedure implies~\cite{MS1} the splitting of a
closed string state into a pair of open strings, or in other words the approach
of a worldsheet bulk operator to the worldsheet boundary. This leads to a
bulk-boundary operator product expansion~\cite{cardy}.

While the recoil operators are unusual and difficult to handle, they have
passed several non-trivial consistency checks. For example, the operators
induce a target space dynamics which is equivalent to that derived from the
Born-Infeld action, even in the case of non-abelian, multiple D-brane
configurations~\cite{MS1}. In this paper we will show that these operators are
also consistent with ${\cal N}=1$ supersymmetry. In particular, we will
construct supersymmetric partners to the recoil operators and show that they
naturally close to a supersymmetric extension of a logarithmic conformal
algebra. The original example of an ${\cal N}=1$ logarithmic superconformal
field theory is provided by a certain class of supersymmetric
Wess-Zumino-Witten models, in which some operator product expansions are found
to contain logarithmic terms~\cite{KM,CKLT}. Supersymmetric extensions of
logarithmic conformal field theories have been subsequently dealt with in
generality
in~\cite{susylog,susylogfrac}. The following construction extends the pair of
impulse operators for D-branes to a consistent deformation of the underlying
supersymmetric worldsheet $\sigma$-model, and at the same time provides one of
the first examples of a logarithmic superconformal field theory.

Besides demonstrating the overall consistency of the recoil operators, the
incorporation of supersymmetric partners enables us to study the quantum
dynamics of the D-brane background in the full setting appropriate to branes in
Type~II superstring theory. We shall uncover a number of remarkable properties
of the ${\cal N}=1$ supermultiplet of recoil operators. For instance, we shall
see that in the case of D0-branes a target space supersymmetrization naturally
leads to the interpretation of the superspace trajectories described by the
impulse operator as those of a certain twisted superparticle~\cite{hull}, with
twist field determined by the angular momentum of the recoiling
D-particles. A certain class of such spinning superparticle theories possesses
the same spectrum as supersymmetric Yang-Mills theory in ten dimensions. The
particular classical configurations can be seen to explicitly
break the target space supersymmetry, which is expected from a time-dependent
D-brane background. This gives a novel physical interpretation of the breaking
of supersymmetry by moving D-branes.

We shall also see that the logarithmic superconformal structure has dramatic
effects on the structure of open string loop amplitudes which describe the
target space quantum corrections to the scattering of elementary string states
off the D-brane. We will find that the fermionic superpartners remove the
logarithmic modular divergences which in the bosonic case arise from string
states associated with logarithmic operators of vanishing conformal
dimension~\cite{KM,MS1}. While the cancellation of instabilities in the purely
bosonic string theory is expected from worldsheet superconformal invariance, in
the logarithmic case it is a non-trivial cancellation because the spinor fields
also exhibit the same sort of logarithmic scaling violations in their
correlation functions. We will see that the stabilization can be attributed to
the identification of the worldsheet scale with an evolution parameter in
target space that was made initially in~\cite{kmw} to extract the proper
logarithmic correlation functions. In the following we will see that this
identification alternatively follows from the requirement of cancellation of
modular divergences and also from the interpretation of the impulse operator in
terms of twisted superparticles with explicitly broken supersymmetry.

The structure of this paper is as follows. In section~2 we describe the ${\cal
N}=1$ supersymmetric extension of the general logarithmic conformal algebra. In
section~3 we explicitly construct the supersymmetric extensions of the impulse
operators and show that they satisfy the structure of a logarithmic
superconformal field theory. In section~4 we show that these same operators
arise naturally from the standard worldsheet supersymmetric Wilson loops which
describe the dynamics of D-branes in the T-dual picture, and also how they
appear canonically in a superspace
formulation of the problem. In section~5 we analyse the interplay between
worldsheet scale transformations and target space time evolution of the
D-brane, and show that the impulse operator explicitly breaks super-Galilean
invariance. We also describe the relationship to the spacetime supersymmetric
Wilson loop and use this to interpret the recoil operator as that of a twisted
superparticle in the case of D0-branes. Finally, in section~6 we demonstrate
the cancellation of modular divergences in superstring annulus amplitudes and
show how it may be most naturally understood in terms of the Zamolodchikov
metric corresponding to the superconformal logarithmic operators.

\newsection{Logarithmic Superconformal Field Theories}

The Virasoro algebra of a two-dimensional conformal field theory is generated
by the worldsheet energy-momentum tensor $T(z)$ with the operator product
expansion
\beq
T(z)\,T(w)=\frac{c/2}{(z-w)^4}+\frac2{(z-w)^2}\,T(w)+\frac1{z-w}\,\partial_w
T(w)+\dots \ ,
\label{TOPE}\eeq
where $c$ is the central charge of the theory, and an ellipsis always denotes
terms in the operator product expansion which are regular as $z\to w$. For a
closed surface these relations are accompanied by their anti-holomorphic
counterparts, while for an open surface the coordinates $z,w$ are real-valued
and parametrize the boundary of the worldsheet. In the following we will be
concerned with the latter case corresponding to open strings and so will not
write any formulas for the anti-holomorphic sector. We shall always set the
worldsheet infrared scale to unity to simplify the formulas which follow.

The simplest logarithmic conformal field theory is characterized by a pair of
operators $C$ and $D$ which become degenerate and span a $2\times2$ Jordan cell
of the Virasoro operators. The two operators then form a logarithmic pair and
their operator product expansion with the energy-momentum tensor involves a
non-trivial mixing~\cite{Gurarie}
\bea
T(z)\,C(w)&=&\frac\Delta{(z-w)^2}\,C(w)+\frac1{z-w}\,\partial_wC(w)+\dots
 \ , \nn\\T(z)\,D(w)&=&\frac\Delta{(z-w)^2}\,D(w)+\frac1{(z-w)^2}\,C(w)
+\frac1{z-w}\,\partial_wD(w)+\dots \ ,
\label{TCD}\eea
where $\Delta$ is the conformal dimension of the operators determined by the
leading logarithmic terms in the conformal blocks of the theory, and an
appropriate normalization of the $D$ operator has been chosen. Because of
(\ref{TCD}), a conformal transformation $z\mapsto w(z)$ mixes the logarithmic
pair as
\beq
\pmatrix{C(z)\cr D(z)\cr}=\left(\frac{\partial w}{\partial z}
\right)^{\pmatrix{\Delta&0\cr1&\Delta\cr}}\,
\pmatrix{C(w)\cr D(w)\cr} \ ,
\label{conftransf}\eeq
from which it follows that their two-point functions are given
by~\cite{Gurarie,CKT}
\bea
\Bigl\langle C(z)\,C(w)\Bigr\rangle&=&0 \ , \nn\\
\Bigl\langle C(z)\,D(w)\Bigr\rangle&=&\frac\xi{(z-w)^{2\Delta}} \ , \nn\\
\Bigl\langle D(z)\,D(w)\Bigr\rangle&=&\frac1{(z-w)^{2\Delta}}\,
\Bigl(-2\xi\ln(z-w)+d\Bigr) \ ,
\label{CD2pt}\eea
where the constant $\xi$ is fixed by the leading logarithmic divergence of the
conformal blocks of the theory and the integration constant $d$ can be changed
by the field redefinition $D\mapsto D+({\rm const.})\,C$. The vanishing of the
$CC$ correlator in (\ref{CD2pt}) is equivalent to the absence of double or
higher logarithmic divergences. From these properties it is evident that the
operator $C$ behaves similarly to an ordinary primary field of scaling
dimension $\Delta$, while the properties of the $D$ operator follow from the
formal identification $D=\partial C/\partial\Delta$.

It is straightforward to write down a superconformal extension of the algebra
(\ref{TCD})~\cite{susylog}. In this paper we shall only deal with ${\cal N}=1$
supersymmetry,
but, as will become apparent, the extension to the ${\cal N}=2$ case is
immediate. The ${\cal N}=1$ superconformal algebra is generated by the
energy-momentum tensor $T(z)$ with the relations (\ref{TOPE}) and an additional
generator, the supercurrent $G(z)$, which is the worldsheet superpartner of
$T(z)$. It has conformal weight $\frac32$ and the operator product expansions
\bea
T(z)\,G(w)&=&\frac{3/2}{(z-w)^2}\,G(w)+\frac1{z-w}\,\partial_wG(w)+\dots \ ,
\nn\\G(z)\,G(w)&=&\frac{\hat c}{(z-w)^3}+\frac2{z-w}\,T(w)+\dots \ ,
\label{TGOPE}\eea
where $\hat c=2c/3$ is the superconformal central charge. We introduce
fermionic fields $\chi^{~}_C$ and $\chi^{~}_D$ which are the worldsheet
superpartners of the operators $C$ and $D$, respectively. For simplicity,
throughout this article,
we
shall work only in the Neveu-Schwarz sector of the theory
corresponding to the choice of anti-periodic boundary conditions on the
worldsheet spinor fields. Then the fields $\chi^{~}_C$ and $\chi^{~}_D$ are
generated by the operator product expansions
\beq
G(z)\,\phi(w)=\frac{1/2}{z-w}\,\chi^{~}_\phi(w)+\dots \ , ~~ \phi=C,D \ .
\label{chiphiOPE}\eeq

One can now derive the ${\cal N}=1$ supersymmetric completion of the
logarithmic conformal algebra (\ref{TCD}). The pair $(C,\chi^{~}_C)$ satisfies
the standard algebraic relations of a primary superconformal multiplet of
dimension $\Delta$, while the additional relations for $\chi^{~}_D$ can be
obtained by differentiating those involving $\chi^{~}_C$ with the formal
identification $\chi^{~}_D=\partial\chi^{~}_C/\partial\Delta$~\cite{susylog}.
The ${\cal N}=1$ logarithmic superconformal algebra is thereby characterized by
the operator product expansions (\ref{TCD}), (\ref{chiphiOPE}), and
\bea
T(z)\,\chi^{~}_C(w)&=&\frac{\Delta+1/2}{(z-w)^2}\,\chi^{~}_C(w)+\frac1{z-w}
\,\partial_w\chi^{~}_C(w)+\dots \ , \nn\\T(z)\,\chi^{~}_D(w)&=&
\frac{\Delta+1/2}{(z-w)^2}\,\chi^{~}_D(w)+\frac1{(z-w)^2}\,\chi^{~}_C(w)+
\frac1{z-w}\,\partial_w\chi^{~}_D(w)+\dots \ , \nn\\G(z)\,\chi^{~}_C(w)&=&
\frac\Delta{(z-w)^2}\,C(w)+\frac{1/2}{z-w}\,\partial_wC(w)+\dots \ ,
\nn\\G(z)\,\chi^{~}_D(w)&=&\frac\Delta{(z-w)^2}\,D(w)+
\frac1{(z-w)^2}\,C(w)+\frac{1/2}{z-w}\,\partial_wD(w)+\dots \ .
\label{SUSYOPE}\eea
In addition to the Green's functions (\ref{CD2pt}), the two-point functions
involving the extra fields can also be readily worked out to be~\cite{susylog}
\bea
\Bigl\langle\phi(z)\,\chi^{~}_{\phi'}(w)\Bigr\rangle&=&0 \ , ~~ \phi,\phi'=
C,D \ , \nn\\\Bigl\langle\chi^{~}_C(z)\,\chi^{~}_C(w)\Bigr
\rangle&=&0 \ ,\nn\\\Bigl\langle\chi^{~}_C(z)\,\chi^{~}_D(w)
\Bigr\rangle&=&\frac{2\Delta\xi}{(z-w)^{2\Delta+1}} \ , \nn
\\\Bigl\langle\chi^{~}_D(z)\,
\chi^{~}_D(w)\Bigr\rangle&=&\frac2{(z-w)^{2\Delta+1}}\,\Bigl(-2\Delta\xi
\ln(z-w)+\xi+\Delta d\Bigr) \ .
\label{SUSY2pt}\eea

Analogous results can be obtained for the three-point correlation functions of
the theory~\cite{susylog,RTAK}. Four-point functions are obtained
in~\cite{flohr1} and for dimension zero fields in~\cite{GabKausch}. For the
cases where there are $n$ degenerate fields which span an $n\times n$ Jordan
cell, the three-point functions have been recently worked out explicitly
in~\cite{flohr2}. It is also possible to
generalize these results to the case where
there is more than one Jordan block. Note that, under the assumption that the
logarithmic partner fields are quasi-primary, any such Jordan block implies the
existence of a Jordan cell for the identity operator, which has vanishing
scaling dimension. Thus, if there exists a Jordan block with $\Delta\neq0$,
then there are automatically at least two Jordan blocks for the logarithmic
conformal field theory.

\newsection{Supersymmetric Impulse Operators for Moving D-Branes}

We will now derive the appropriate supersymmetric vertex operator describing
the recoil of a D-brane and show that it naturally satisfies a logarithmic
superconformal algebra. For simplicity, we shall deal only with the case of
0-branes, but the extensions to generic $p$-branes are straightforward. At
tree-level in open string perturbation theory, such a configuration is
described by the harmonic string coordinates $x^\mu=(x^0,x^i)$ which map a disc
$\Sigma$ into flat 1+9 dimensional spacetime with the Dirichlet boundary
conditions $x^i|_{\partial\Sigma}=0$ along the transverse directions to the
brane and Neumann ones $\partial^{~}_{\!\perp} x^0|_{\partial\Sigma}=0$ along
the D0-brane worldline. Here $\partial^{~}_{\!\perp}$ denotes the normal
derivative at the boundary of the string worldsheet $\Sigma$ which is
parametrized by the periodic coordinate $\tau\in[0,1]$. The bulk of $\Sigma$ is
parametrized by coordinates $\sigma^\alpha$, $\alpha=1,2$. The bosonic vertex
operator describing the motion of the brane is given by~\cite{Dmoving}
\bea
V_{\rm D}^{\rm bos}&=&\exp\left(-\frac1{2\pi\alpha'}\,\int\limits_\Sigma
d^2\sigma~\eta^{\alpha\beta}\,\partial_\alpha\left[Y_i\Bigl(x^0(\sigma)\Bigr)
\,\partial_\beta
x^i(\sigma)\right]\right)\nn\\&=&\exp\left(-\frac1{2\pi\alpha'}\,
\int\limits_0^1d\tau~Y_i\Bigl(x^0(\tau)\Bigr)\,\partial^{~}_{\!\perp} x^i(
\tau)\right) \ ,
\label{vertexD}\eea
where $\alpha'$ is the string slope,
$\partial_\alpha=\partial/\partial\sigma^\alpha$, and
$Y_i(x^0)=\delta_{ij}\,Y^j(x^0)$ describes the trajectory of the D0-brane as it
moves in spacetime.

The recoil of a heavy D-brane due to the scattering of closed string states may
be described in an impulse approximation by inserting appropriate factors of
the usual Heaviside function $\Theta(x^0)$ into (\ref{vertexD}). This describes
a non-relativistic 0-brane which begins moving at time $x^0=0$ from the initial
position $y_i$ with a constant velocity $u_i$. The appropriate trajectory is
given by the operator~\cite{kmw}
\beq
Y_i(x^0)=y_i\,C_{\epsilon}(x^0)+u_i\,D_{\epsilon}(x^0) \ ,
\label{Yirecoil}\eeq
where we have introduced the operators
\beq
C_\epsilon(x^0)=\alpha'\,\epsilon\,\Theta_\epsilon(x^0) \ , ~~
D_\epsilon(x^0)=x^0\,\Theta_\epsilon(x^0) \ ,
\label{CDops}\eeq
with $\Theta_\epsilon(x^0)$ the regulated step function which is defined by the
Fourier integral transformation
\beq
\Theta_\epsilon(x^0)=\frac1{2\pi i}\,\int\limits_{-\infty}^\infty\frac{d\omega}
{\omega-i\epsilon}~\e^{i\omega x^0} \ .
\label{ThetaFourier}\eeq
This integral representation is needed to make the Heaviside function
well-defined as an {\it operator}. In the limit $\epsilon\to0^+$, it reduces
via the residue theorem to the usual step function. The operator
$C_\epsilon(x^0)$ is required in (\ref{Yirecoil}) by scale invariance. Note
that the center of mass coordinate $y_i$ appears with a factor of
$\epsilon\to0^+$, so that the first operator in (\ref{Yirecoil}) represents a
small uncertainty in the initial position of the D-brane induced by stringy
effects~\cite{kmw}. The pair of fields (\ref{CDops}) are interpreted as
functions of the coordinate $z$ on the upper complex half-plane, which is
identified with the boundary variable $\tau$ in (\ref{vertexD}). This
interpretation is possible because the boundary vertex operator (\ref{vertexD})
is a total derivative and so can be thought of as a bulk deformation of the
underlying free bosonic conformal $\sigma$-model on $\Sigma$ (in the conformal
gauge). The impulse operator (\ref{vertexD},\ref{Yirecoil}) then describes the
appropriate change of state of the D-brane background because it has
non-vanishing matrix elements between different string states. It can be
thought of as generating the action of the Poincar\'e group on the 0-brane,
with $y_i$ parametrizing translations and $u_i$ parametrizing boosts in the
transverse directions.

By using the representation (\ref{ThetaFourier}) and the fact that the tachyon
vertex operator $\e^{i\omega x^0}$ has conformal dimension $\alpha'\omega^2/2$,
it can be shown~\cite{kmw} that the operators (\ref{CDops}) form a degenerate
pair which generate a logarithmic conformal algebra (\ref{TCD}) with conformal
dimension $\Delta=\Delta_\epsilon$, where
\beq
\Delta_\epsilon=-\frac{\alpha'\epsilon^2}2 \ .
\label{Deltavarep}\eeq
The total dimension of the impulse operator (\ref{vertexD},\ref{Yirecoil}) is
$h_\epsilon=1+\Delta_\epsilon$, and so for $\epsilon\neq0$ it describes a
relevant deformation of the underlying worldsheet conformal $\sigma$-model. The
existence of such a deformation implies that the resulting string theory is
slightly non-critical and leads to the change of state of the D-brane
background.

The two-point functions of the operators (\ref{CDops}) can be computed
explicitly to be~\cite{kmw}
\bea
\Bigl\langle C_\epsilon(z)\,C_\epsilon(w)\Bigr\rangle&=&\frac1{4\pi}\,
\sqrt{\frac{(\alpha')^3}{\epsilon^2\ln\Lambda}}\,\left[\frac{\sqrt\pi}2~
{}^{~}_1F^{~}_1\Bigl(\mbox{$\frac12$}\,,\,\mbox{$\frac12$}\,;4\epsilon^2\alpha'
\ln(z-w)\Bigr)\right.\nn\\&&-\left.2\,\sqrt{\epsilon^2\alpha'\ln(z-w)}~
{}^{~}_1F^{~}_1\Bigl(1\,,\,\mbox{$\frac32$}\,;4\epsilon^2\alpha'\ln(z-w)\Bigr)
\right] \ , \nn\\\Bigl\langle C_\epsilon(z)\,D_\epsilon(w)\Bigr\rangle&=&
\frac1{4\pi\epsilon^3}\,\sqrt{\frac1{2\alpha'\ln\Lambda}}\,\left[
\frac{\sqrt\pi}8~{}^{~}_1F^{~}_1\Bigl(\mbox{$\frac12$}\,,\,-\mbox{$\frac12$}\,;
4\epsilon^2\alpha'\ln(z-w)\Bigr)\right.\nn\\&&+\left.\frac{16}3\,\Bigl(
\epsilon^2\alpha'\ln(z-w)\Bigr)^{3/2}~{}^{~}_1F^{~}_1\Bigl(2\,,\,
\mbox{$\frac52$}\,;4\epsilon^2\alpha'\ln(z-w)\Bigr)\right] \ , \nn\\
\Bigl\langle D_\epsilon(z)\,D_\epsilon(w)\Bigr\rangle&=&
\frac1{\epsilon^2\alpha'}
\,\Bigl\langle C_\epsilon(z)\,D_\epsilon(w)\Bigr\rangle \ ,
\label{CDepsilon2pt}\eea
where $\Lambda\to0$ is the worldsheet ultraviolet cutoff which arises from the
short-distance propagator
\beq
\lim_{z\to w}\,\Bigl\langle x^0(z)\,x^0(w)\Bigr\rangle=-2\alpha'\ln\Lambda \ .
\label{shortdistprop}\eeq
Here we have used the standard bulk Green's function in the upper half-plane,
as the effects of worldsheet boundaries will not be relevant for the ensuing
analysis.\footnote{\baselineskip=12pt Boundary effects in logarithmic conformal
field theories have been analysed in \cite{MS1,kw}.} This is again justified by
the bulk form of the vertex operator (\ref{vertexD}), and indeed it can be
shown that using the full expression for the propagator on the disc does not
alter any results~\cite{MS1}. It is then straightforward to see~\cite{kmw} that
in the correlated limit $\epsilon,\Lambda\to0^+$, with
\beq
\frac1{\epsilon^2}=-2\alpha'\ln\Lambda \ ,
\label{epsilonLambdarel}\eeq
the correlators (\ref{CDepsilon2pt}) reduce at order $\epsilon^2$ to the
canonical two-point correlation functions (\ref{CD2pt}) of the logarithmic
conformal algebra, with conformal dimension (\ref{Deltavarep}) and the
normalization constants
\beq
\xi=\frac{\pi^{3/2}\,\alpha'}2 \ , ~~ d=d_\epsilon
=\frac{\pi^{3/2}}{2\epsilon^2} \ .
\label{aconstsrecoil}\eeq
Note that the singular behaviour of the constant $d_\epsilon$ in
(\ref{aconstsrecoil}) is not harmful, because it can be removed by considering
instead the connected correlation functions of the theory~\cite{kmw}.

We will now derive the ${\cal N}=1$ supersymmetric completion of the impulse
operator (\ref{vertexD},\ref{Yirecoil}). For this, we introduce $2\times2$
Dirac matrices $\rho^\alpha$, $\alpha=1,2$, and real two-component Majorana
fermion fields $\psi^\mu$ which are the worldsheet superpartners of the string
embedding fields $x^\mu$. A convenient basis for the worldsheet spinors is
given by
\beq
\rho^1=\pmatrix{0&-i\cr i&0\cr} \ , ~~ \rho^2=\pmatrix{0& i\cr i&0\cr} \ ,
\label{rhobasis}\eeq
in which the fermion fields decompose as
\beq
\psi^\mu=\pmatrix{\psi^\mu_-\cr\psi^\mu_+\cr} \ .
\label{WSspinordecomp}\eeq
The fields (\ref{WSspinordecomp}) obey the boundary conditions
$\psi_+^\mu|_{\partial\Sigma}=\pm\,\psi_-^\mu|_{\partial\Sigma}$, where the
sign depends on whether they belong to the Ramond or Neveu-Schwarz
sector of the worldsheet theory~\cite{GSW}. The global worldsheet supersymmetry
is determined by the supercharge $\cal Q$ which generates the infinitesimal
${\cal N}=1$ supersymmetry transformations~\cite{GSW}
\bea
\Bigl[{\cal Q}\,,\,x^\mu\Bigr]&=&\psi^\mu \ , \non\Bigl\{{\cal Q}\,,\,
\psi^\mu\Bigr\}&=&-i\,\rho^\alpha\,\partial_\alpha x^\mu \ .
\label{WSsusytransfs}\eea

The fermionic fields $\psi_+(z)$ have conformal dimension $\frac12$, and from
(\ref{WSsusytransfs}) it follows that the superpartner of the tachyon vertex
operator $\e^{i\omega x^0}$ is $\sqrt{\alpha'}\,\omega\,\psi_+^0\,\e^{i\omega
x^0}$, so that in the Neveu-Schwarz sector we may write
\beq
G(z)~\e^{i\omega x^0(w)}=\frac{\sqrt{\alpha'}\,\omega/2}{z-w}\,
\psi_+^0(w)~\e^{i\omega x^0(w)}+\dots \ .
\label{tachyonsusy}\eeq
In what follows it will be important to note the factor of
$\sqrt{\alpha'}\,\omega$ that appears in the supersymmetry transformation
(\ref{tachyonsusy}). Because of it, and the fact that the tachyon vertex
operator has conformal dimension $\alpha'\omega^2/2$, the inverse
transformation is given by
\beq
G(z)\,\psi_+^0(w)~\e^{i\omega x^0(w)}=\frac{\sqrt{\alpha'}\,\omega/2}{(z-w)^2}~
\e^{i\omega x^0(w)}+\frac{i/2\sqrt{\alpha'}}{z-w}\,
\Bigl(\partial_wx^0(w)\Bigl)~\e^{i\omega x^0(w)}+\dots \ .
\label{tachyonsusyinv}\eeq

To compute the superpartners of the logarithmic pair (\ref{CDops}), we use
(\ref{ThetaFourier}) and (\ref{tachyonsusy}) to write
\beq
G(z)\,C_\epsilon(w)=\frac{\epsilon\,(\alpha')^{3/2}/4\pi i}{z-w}\,
\psi_+^0(w)\,\int\limits_{-\infty}^\infty\frac{d\omega}{\omega-i\epsilon}~
\Bigl[(\omega-i\epsilon)+i\epsilon\Bigr]~\e^{i\omega x^0(w)}+\dots \ .
\label{GCepsilon}\eeq
In the first term of the integrand in (\ref{GCepsilon}) there is no pole and so
after contour integration it vanishes. Formally it is a delta functional
$\delta(x^0(w))$ which we neglect since we are interested here in only the
asymptotic time-dependence of string solitons. Then, only the second term
contributes, and comparing with (\ref{chiphiOPE}) we find
\beq
\chi^{~}_{C_\epsilon}(x^0,\psi^0)=i\,\epsilon\,C_\epsilon(x^0)\,\psi_+^0 \ .
\label{chiCepsilon}\eeq
Similarly, we have
\beq
G(z)\,D_\epsilon(w)=-\frac{\sqrt{\alpha'}/4\pi}{z-w}\,\psi_+^0(w)\,
\int\limits_{-\infty}^\infty\frac{d\omega}{(\omega-i\epsilon)^2}~
\Bigl[(\omega-i\epsilon)+i\epsilon\Bigr]~\e^{i\omega x^0(w)}+\dots \ ,
\label{GDepsilon}\eeq
which using (\ref{chiphiOPE}) gives
\beq
\chi^{~}_{D_\epsilon}(x^0,\psi^0)=i\,\left(\epsilon\,D_\epsilon(x^0)-\frac1
{\epsilon\,\alpha'}\,C_\epsilon(x^0)\right)\,\psi_+^0 \ .
\label{chiDepsilon}\eeq
The operators (\ref{chiCepsilon}) and (\ref{chiDepsilon}) have conformal
dimension $\Delta_\epsilon+\frac12$.

It is straightforward to now check that the remaining relations of the ${\cal
N}=1$ logarithmic superconformal algebra are satisfied. By using (\ref{TCD}),
(\ref{Deltavarep}), (\ref{chiCepsilon}) and (\ref{chiDepsilon}), it is easy to
verify the first two operator product expansions of (\ref{SUSYOPE}) in this
case. For the operator products with the fermionic supercurrent, we use in
addition the Fourier integral (\ref{ThetaFourier}) along with
(\ref{tachyonsusyinv}) to get
\bea
G(z)\,\chi^{~}_{C_\epsilon}(w)&=&-\frac{\epsilon^2(\alpha')^{3/2}/4\pi i}
{(z-w)^2}\,\int\limits_{-\infty}^\infty\frac{d\omega}{\omega-i\epsilon}~
\Bigl[(\omega-i\epsilon)+i\epsilon\Bigr]\e^{i\omega x^0(w)}\nn\\&&
+\,\frac1{z-w}\,\partial_w\chi^{~}_{C_\epsilon}(w)+\dots\nn\\
&=&-\frac{\sqrt{\alpha'}\,\epsilon^2/2}{(z-w)^2}\,C_\epsilon(w)+\frac1{z-w}\,
\partial_w\chi^{~}_{C_\epsilon}(w)+\dots \ , \\&&~~~~~\nn\\
G(z)\,\chi^{~}_{D_\epsilon}(w)&=&
-\frac{\sqrt{\alpha'}/4\pi}{(z-w)^2}\,\int\limits_{-\infty}^\infty
\frac{d\omega}{\omega-i\epsilon}\,\left[\Bigl((\omega-i\epsilon)+i\epsilon
\Bigr)+\frac{i\epsilon}{\omega-i\epsilon}\,\Bigl((\omega-i\epsilon)+i
\epsilon\Bigr)\right]\nn\\&&\times\,\e^{i\omega x^0(w)}+\frac1{z-w}\,
\partial_w\chi^{~}_{D_\epsilon}(w)+\dots\nn\\&=&\frac{1/\sqrt{\alpha'}}
{(z-w)^2}\,\left(C_\epsilon(w)-\frac{\alpha'\epsilon^2}2\,D_\epsilon(w)\right)
+\frac1{z-w}\,\partial_w\chi^{~}_{D_\epsilon}(w)+\dots \ ,
\label{Gchiepsilon}\eeq
which also agree with (\ref{SUSYOPE}) in this case.

For the two-point correlation functions (\ref{SUSY2pt}), we use the fermionic
Green's function in the upper half-plane,
\beq
\Bigl\langle\psi_+^0(z)\,\psi_+^0(w)\Bigr\rangle=\frac1{z-w} \ ,
\label{fermGF}\eeq
and the fact that bosonic and fermionic field correlators factorize from each
other in the free superconformal $\sigma$-model on $\Sigma$. The first set of
relations in (\ref{SUSY2pt}) are then satisfied in this case because
$\langle\psi_+^0(z)\rangle=0$. The second relation holds to order $\epsilon^4$
since $\Delta_\epsilon\propto\epsilon^2$ and $\langle
C_\epsilon(z)C_\epsilon(w)\rangle=0$ to order $\epsilon^2$. For the remaining
correlators, we use (\ref{CD2pt}), (\ref{chiCepsilon}), (\ref{chiDepsilon}),
(\ref{fermGF}) and factorization to compute
\bea
\Bigl\langle\chi^{~}_{C_\epsilon}(z)\,\chi^{~}_{D_\epsilon}(w)\Bigr\rangle&=&
-\frac{\epsilon^2\xi}{(z-w)^{2\Delta_\epsilon+1}} \ , \nn\\
\Bigl\langle\chi^{~}_{D_\epsilon}(z)\,\chi^{~}_{D_\epsilon}(w)\Bigr\rangle&=&
\frac1{(z-w)^{2\Delta_\epsilon+1}}\,\left[\frac{2\xi}{\alpha'}-
\epsilon^2\Bigl(-2\xi\ln(z-w)+d_\epsilon\Bigr)\right] \ ,
\label{chiepsilon2pt}\eea
which upon using (\ref{Deltavarep}) are also seen to agree with
(\ref{SUSY2pt}). Therefore, the supersymmetric extensions (\ref{chiCepsilon})
and (\ref{chiDepsilon}) of the impulse operators (\ref{CDops}) give precisely
the right combinations of operators that generate the full algebraic structure
of a logarithmic superconformal field theory. This yields a non-trivial
realization of the supersymmetric completion of the previous section, and
illustrates the overall consistency of the impulse operators describing the
dynamics of D-branes in closed string scattering states.

\newsection{Superspace Formalism}

We will now derive the explicit form of the supersymmetric extension of the
impulse vertex operator (\ref{vertexD},\ref{Yirecoil}). For this, we consider
the Wilson loop operator
\beq
W[A]=\exp i\oint\limits_{\partial\Sigma}A_\mu(x)~dx^\mu=\exp i\int\limits_0^1
d\tau~\dot x^\mu(\tau)\,A_\mu\Bigl(x(\tau)\Bigr) \ ,
\label{Wilsonloop}\eeq
where $A_\mu$ is a $U(1)$ gauge field in ten dimensions, and $\dot
x^\mu(\tau)=dx^\mu(\tau)/d\tau$. T-duality maps the operator (\ref{Wilsonloop})
onto the vertex operator (\ref{vertexD}) for a moving D-brane by the rule
$\partial_\alpha x^i\mapsto
i\,\eta^{\beta\gamma}\,\epsilon_{\alpha\beta}\,\partial_\gamma x^i$ and the
resulting replacement of Neumann boundary conditions for $x^i$ with Dirichlet
ones~\cite{Bachas}. The spatial components of the Chan-Paton gauge field map
onto the brane trajectory as $A_i=Y_i/2\pi\alpha'$, while the temporal
component $A_0$ becomes a $U(1)$ gauge field on the D-particle worldline.

The minimal ${\cal N}=1$ worldsheet supersymmetric extension of the operator
(\ref{Wilsonloop}) is given by
\beq
{\cal W}[A,\psi]=W[A]\,\exp\left(-\frac12\,\int\limits_0^1d\tau~
F_{\mu\nu}\,\overline{\psi}^{\,\mu}\,\rho^1\,\psi^\nu\right) \ ,
\label{WSsusyWilson}\eeq
where $F_{\mu\nu}$ is the corresponding gauge field strength tensor. For the
recoil trajectory (\ref{Yirecoil}), an elementary computation using the contour
integration techniques outlined in the previous section gives $F_{ij}=0$ and
\beq
F_{0i}(x^0)=\frac{\delta A_i(x^0)}{\delta x^0}=\frac i{2\pi\alpha'}\,\left[
y_i\,\epsilon\,C_\epsilon(x^0)+u_i\left(\epsilon\,D_\epsilon(x^0)-
\frac1{\epsilon\,\alpha'}\,C_\epsilon(x^0)\right)\right] \ .
\label{F0i}\eeq
This shows that, in the T-dual Neumann picture, the canonical supersymmetric
extension of the $U(1)$ Wilson loop operator (\ref{WSsusyWilson}) yields {\it
precisely} the couplings to the operators $\chi^{~}_{C_\epsilon}$ and
$\chi^{~}_{D_\epsilon}$ that were computed in the previous section from the
supersymmetric completion of the worldsheet logarithmic conformal algebra.

T-duality acts on the worldsheet fermion fields (\ref{WSspinordecomp}) by
reversing the sign of their right-moving components $\psi_-^\mu$. By using
(\ref{WSsusyWilson},\ref{F0i}) we may thereby write down the supersymmetric
extension of the impulse operator for moving D0-branes,
\bea
V_{\rm D}^{\rm susy}&=&\exp\left(-\frac1{2\pi\alpha'}\,\int\limits_0^1d\tau~
\left\{\left[y_i\,C_\epsilon\Bigl(x^0(\tau)\Bigr)+u_i\,D_\epsilon
\Bigl(x^0(\tau)\Bigr)\right]\,\partial^{~}_{\!\perp} x^i(\tau)\right.\right.
\nn\\&&+\Biggl.\left.\left[
y_i\,\chi_{C_\epsilon}^{~}\Bigl(x^0(\tau)\,,\,\psi^0(\tau)\Bigr)+u_i\,
\chi^{~}_{D_\epsilon}\Bigl(x^0(\tau)\,,\,\psi^0(\tau)\Bigr)\right]\,
\psi^i(\tau)\right\}\Biggr) \ ,
\label{vertexDsusy}\eea
where we have dropped the $\pm$ subscripts on the fermion fields in
(\ref{WSspinordecomp}), and the logarithmic superconformal operators in
(\ref{vertexDsusy}) are given by (\ref{CDops}), (\ref{chiCepsilon}) and
(\ref{chiDepsilon}). The vertex operator (\ref{vertexDsusy}) can be expressed
in a more compact form which makes its supersymmetry manifest. For this, we
extend the disc $\Sigma$ to an ${\cal N}=1$ super-Riemann surface $\hat\Sigma$
with coordinates $(Z,\bar
Z)=(z,\theta,\bar z,\bar\theta\,)$, where $\theta$ is a complex Grassmann
variable, and with corresponding superspace covariant derivatives ${\cal
D}_Z=\partial_\theta+\theta\,\partial_z$. Given a bosonic field $\phi(z)$ with
superpartner $\chi^{~}_\phi(z)$, we introduce the chiral worldsheet superfields
\beq
\Phi_\phi(z,\theta)=\phi(z)+\theta\,\chi^{~}_\phi(z) \ ,
\label{Phisuper}\eeq
and correspondingly we make the embedding space of the superstring an ${\cal
N}=1$ superspace with chiral scalar superfields
$X^i(z,\theta)=x^i(z)+\theta\,\psi^i(z)$. Then the impulse operator
(\ref{vertexDsusy}) can be written in a manifestly supersymmetric form in terms
of superspace quantities as
\beq
V_{\rm D}^{\rm susy}=\exp\left[-\frac1{2\pi\alpha'}\,\oint
\limits_{\partial\hat\Sigma}d\tau~d\theta~
\Bigl(y_i\,\Phi_{C_\epsilon}(\tau,\theta)+u_i\,\Phi_{D_\epsilon}(\tau,\theta)
\Bigr)\,{\cal D}^{~}_{\!\perp}X^i(\tau,\theta)\right] \ ,
\label{VDsusysuperspace}\eeq
where in (\ref{VDsusysuperspace}) the Grassmann coordinate $\theta$ is real.

In fact, the algebraic relations of the logarithmic superconformal algebra can
be most elegantly expressed in superspace notation. For this, we introduce the
super-stress tensor ${\cal T}(Z)=G(z)+\theta\,T(z)$, and define the quantities
$Z_{12}=z_1-z_2-\theta_1\theta_2$ and $\theta_{12}=\theta_1-\theta_2$
corresponding to a pair of holomorphic superspace coordinates
$Z_1=(z_1,\theta_1)$ and $Z_2=(z_2,\theta_2)$. Then the operator product
expansions (\ref{TOPE}), (\ref{TCD}) and (\ref{TGOPE})--(\ref{SUSYOPE}) can
also be written in terms of superspace quantities as
\bea
{\cal T}(Z_1)\,{\cal T}(Z_2)&=&\frac{\hat c/4}{(Z_{12})^3}+\frac{3\theta_{12}
/2}{(Z_{12})^2}\,{\cal T}(Z_2)+\frac{1/2}{Z_{12}}\,{\cal D}_{Z_2}{\cal T}(Z_2)
+\frac{\theta_{12}}{Z_{12}}\,\partial_{z_2}{\cal T}(Z_2)+\dots \ , \nn\\
{\cal T}(Z_1)\,\Phi_C(Z_2)&=&\frac{\theta_{12}\,\Delta/2}{(Z_{12})^2}\,
\Phi_C(Z_2)+\frac{1/2}{Z_{12}}\,{\cal D}_{Z_2}\Phi_C(Z_2)+\frac{\theta_{12}}
{Z_{12}}\,\partial_{z_2}\Phi_C(Z_2)+\dots \ , \nn\\
{\cal T}(Z_1)\,\Phi_D(Z_2)&=&\frac{\theta_{12}\,\Delta/2}{(Z_{12})^2}\,
\Phi_D(Z_2)+\frac{\theta_{12}/2}{(Z_{12})^2}\,\Phi_C(Z_2)\nn\\&&
+\,\frac{1/2}{Z_{12}}\,{\cal D}_{Z_2}\Phi_D(Z_2)+\frac{\theta_{12}}
{Z_{12}}\,\partial_{z_2}\Phi_D(Z_2)+\dots \ ,
\label{superspaceOPE}\eea
while the two-point functions (\ref{SUSY2pt}) may be expressed as
\bea
\Bigl\langle\Phi_C(Z_1)\,\Phi_C(Z_2)\Bigr\rangle&=&0 \ , \nn\\
\Bigl\langle\Phi_C(Z_1)\,\Phi_D(Z_2)\Bigr\rangle&=&\frac\xi{(Z_{12})^{2
\Delta}} \ , \nn\\\Bigl\langle\Phi_D(Z_1)\,\Phi_D(Z_2)\Bigr\rangle&=&
\frac1{(Z_{12})^{2\Delta}}\,\Bigl(-2\xi\ln Z_{12}+d\Bigr) \ .
\label{superspace2pt}\eea
This superspace formalism also generalizes to the construction of higher-order
correlation functions which are built from appropriate coordinate invariants of
the supergroup $OSp(1|2)$~\cite{susylog}. It emphasizes how the impulse
operator (\ref{VDsusysuperspace}), and the ensuing logarithmic algebra
(\ref{superspaceOPE},\ref{superspace2pt}), is the natural supersymmetrization
of the recoil operators for D-branes.

\newsection{Target Space Formalism}

Let us now describe the target space properties of the logarithmic
superconformal algebra that we have derived. A worldsheet finite-size scaling
\beq
\Lambda~\longmapsto~\Lambda'=\Lambda~\e^{-t/\sqrt{\alpha'}}
\label{Lambdascale}\eeq
induces from (\ref{epsilonLambdarel}) a transformation of the target space
regularization parameter,
\beq
\epsilon~\longmapsto~\epsilon'=\epsilon+\epsilon^3\,t\,\sqrt{\alpha'}
+O(\epsilon^5) \ .
\label{epsilonscale}\eeq
By using (\ref{aconstsrecoil}) and the ensuing scale dependence of the
correlation functions (\ref{CD2pt}) we may then infer the transformation rules
\beq
C_{\epsilon'}=C_\epsilon \ , ~~ D_{\epsilon'}=D_\epsilon-\frac
t{\sqrt{\alpha'}}\,C_\epsilon
\label{CDscale}\eeq
to order $\epsilon^2$. It follows that, in order to maintain conformal
invariance, the $\sigma$-model coupling constants in (\ref{vertexDsusy}) must
transform as $y_i\mapsto y_i+(t/\sqrt{\alpha'}\,)\,u_i$, $u_i\mapsto u_i$, and
thus a worldsheet scale transformation leads to a Galilean boost of the D-brane
in target space.

However, by using (\ref{chiCepsilon}), (\ref{chiDepsilon}),
(\ref{epsilonscale}) and (\ref{CDscale}), we see that the superconformal
partners of the logarithmic operators are scale-invariant to order
$\epsilon^2$,
\beq
\chi^{~}_{C_{\epsilon'}}=\chi^{~}_{C_\epsilon} \ , ~~ \chi^{~}_{D_{\epsilon'}}=
\chi^{~}_{D_\epsilon} \ .
\label{chiCDscale}\eeq
The invariance property (\ref{chiCDscale}) can also be deduced from the scale
independence to order $\epsilon^2$ of the two-point correlators
(\ref{SUSY2pt}), in which the scale dependent constant $d_\epsilon$ appears
only in the invariant combination $\Delta_\epsilon d_\epsilon\sim
O(\epsilon^0)$. This means that the operator (\ref{vertexDsusy}) describes the
evolution of the D0-brane in target space with respect to only the ordinary,
bosonic Galilean group. In other words, if we introduce a superspace and
worldsheet superfields as in (\ref{Phisuper}), then a worldsheet scale
transformation in the present case acts only on the bosonic part of the
superspace. This property is of course very particular to the explicit scale
dependence of the recoil
superpartners (\ref{chiDepsilon}) in the logarithmic superconformal algebra.

The fact that the super-Galilean group is not represented in the
non-relativistic dynamics of D-branes is merely a reflection of the fact that
the motion of the brane explicitly breaks target space supersymmetry. Indeed,
while the deformed $\sigma$-model that we have been working with possesses
${\cal N}=1$ {\it worldsheet} supersymmetry, it is only after the appropriate
sum over worldsheet spin structures and the GSO projection that it has the
possibility of possessing spacetime supersymmetry. To understand better the
breaking of target space supersymmetry within the present formalism, we now
appeal to an explicit spacetime supersymmetrization of the Wilson loop operator
(\ref{Wilsonloop}). This will produce a Green-Schwarz representation of the
spacetime
supersymmetric impulse operator in the dual Neumann picture, and also yield a
physical interpretation of the supersymmetric vertex operator
(\ref{vertexDsusy}).

For this, we regard the Chan-Paton gauge field $A_\mu$ as the first component
of the ten-dimensional ${\cal N}=1$ Maxwell supermultiplet. Its superpartner is
therefore a Majorana-Weyl fermion field $\lambda$ with 32 real components. We
introduce Dirac matrices $\Gamma_\mu$ in 1+9 dimensions, and define
$\Gamma_{\mu\nu}=\frac12\,[\Gamma_\mu,\Gamma_\nu]$. The loop parametrization
$x^\mu(\tau)$ has superpartner $\vartheta(\tau)$ which couples to the photino
field $\lambda$. Then, the spacetime supersymmetric extension of
(\ref{Wilsonloop}) is given by the finite supersymmetry transformation
\beq
{\sf W}[A,\lambda]=\exp\left(\,\int\limits_0^1d\tau~
\overline{\vartheta}(\tau)\,{\sf Q}
\right)~W[A]~\exp\left(-\int\limits_0^1d\tau~\overline{\vartheta}(\tau)
\,{\sf Q}\right) \ ,
\label{susyWilsonloop}\eeq
where the supercharge $\sf Q$ generates the infinitesimal ${\cal N}=1$
supersymmetry transformations~\cite{GSW}
\bea
\Bigl[{\sf Q}\,,\,A_\mu\Bigr]&=&\frac i2\,\Gamma_\mu\,\lambda \ , \non
\Bigl\{{\sf Q}\,,\,\lambda\Bigr\}&=&-\frac14\,
\Gamma_{\mu\nu}\,F^{\mu\nu} \ , \nonumber\\ \Bigl[{\sf Q}\,,\,x^\mu\Bigr]
&=&\frac i4\,\Gamma^\mu\,\vartheta \ , \non \Bigl\{{\sf Q}
\,,\,\vartheta\Bigr\}&=&4 \ .
\label{susyspacetime}\eeq
By using the Baker-Campbell-Hausdorff formula, the supersymmetric Wilson loop
(\ref{susyWilsonloop}) thereby admits an expansion
\bea
{\sf W}[A,\lambda]&=&\exp i\int\limits_0^1d\tau~\left(\dot x^\mu\,A_\mu
+\frac i4\,A_\mu\,\overline{\vartheta}\,\Gamma^\mu\,\dot\vartheta\right.\non&&
+\left.\frac i2\,\dot x^\mu\,\overline{\vartheta}\,\Gamma_\mu\,\lambda
+\frac i{16}\,\dot x^\mu F^{\nu\lambda}\,\overline{\vartheta}\,\Gamma_\mu\,
\Gamma_{\nu\lambda}\,\vartheta+\dots\right) \ ,
\label{susyWilsonexp}\eea
where the ellipsis in (\ref{susyWilsonexp}) denotes contributions from
higher-order fluctuation modes of the fields.

To identify the ten-dimensional supermultiplet which is T-dual to the
worldsheet recoil supermultiplet of (\ref{vertexDsusy}), we use the
supersymmetry algebra (\ref{susyspacetime}) to get
$\Gamma_i\,\lambda=\frac12\,F_{0i}(x^0)\,\Gamma^0\,\vartheta$, with
$F_{0i}(x^0)$ given by (\ref{F0i}). We then find that the target space
supermultiplet describing the recoil of a D0-brane is given by the
dimensionally reduced supersymmetric Yang-Mills fields
\bea
A_i(x^0)&=&\frac1{2\pi\alpha'}\,\Bigl(y_i\,C_\epsilon(x^0)+u_i\,D_\epsilon
(x^0)\Bigr) \ , \nn\\\lambda(x^0,\vartheta)&=&\frac1{36\pi\alpha'}\,\Gamma^i\,
\Bigl(y_i\,\chi^{~}_{C_\epsilon}(x^0,\Gamma^0\,\vartheta)+u_i\,
\chi^{~}_{D_\epsilon}(x^0,\Gamma^0\,\vartheta)\Bigr) \ .
\label{recoilsupermult}\eea
Therefore, the logarithmic superconformal partners to the basic recoil
operators also arise naturally in the T-dual Green-Schwarz formalism.

By using (\ref{susyWilsonexp}) and (\ref{recoilsupermult}) we can now lend a
physical interpretation to the supersymmetric impulse operator. For simplicity,
we shall neglect the stringy fluctuations in the center of mass coordinates of
the D-brane and take $y_i=0$. We consider only the long-time dynamics of the
string soliton, i.e. we take $x^0>0$ and effectively set the Heaviside function
$\Theta_\epsilon(x^0)$ to unity everywhere. We will also choose the gauge
$A_0(x)=0$. The bosonic part of the Maxwell supermultiplet of course describes
the free, non-relativistic geodesic motion of the D0-brane in flat space. To
see what sort of particle kinematics is represented by the full supermultiplet,
we substitute $A_i=u_i\,x^0/2\pi\alpha'$ and
$\lambda=u\slash\,\Gamma^0\,\vartheta/36\pi\alpha'$ into (\ref{susyWilsonexp}),
where $u\slash=u_i\,\Gamma^i$, and we have again ignored stringy $O(\epsilon)$
uncertainties in position and velocity. Note that, generally, the fermionic
operator (\ref{chiDepsilon}) also induces a velocity-dependent stringy
contribution to the phase space uncertainty principle in the sense described
in~\cite{kmw}. This is reminescent of the energy-dependent smearings that were
found in~\cite{MS1}. Heuristically, this identical stringy smearing of position
and velocity is responsible for the violation of super-Galilean invariance in
(\ref{chiCDscale}).

With these substitutions we find ${\sf W}[A,\lambda]=\e^{i\,{\sf
S}/2\pi\alpha'}$, where
\bea
{\sf S}&=&\int\limits_0^1d\tau~\left(\dot x^i\,u_i\,x^0+\frac i4\,x^0\,
\overline{\vartheta}\,u\slash\,\dot\vartheta+\frac i{36}\,\dot x^0\,
\overline{\vartheta}\,u\slash\,\vartheta-\frac i4\,\dot x^i\,u_i\,
\overline{\vartheta}\,\Gamma^0\,\vartheta\right.\nn\\&&+\left.
\frac i{32}\,\dot x^0\,\overline{\vartheta}\,\left[\Gamma^0\,,\,u\slash
\right]\,\vartheta+\frac i{32}\,\dot{x\slash}\,\overline{\vartheta}
\left[\Gamma^0\,,\,u\slash\right]\,\vartheta+\dots\right)
\label{superpartaction}\eea
can be interpreted as the action of a certain kind of superparticle in the
${\cal N}=1$ superspace spanned by the coordinates
$(x^i,\vartheta,\overline{\vartheta}\,)$ and with worldline parametrized by the
loop coordinate $\tau$. To identify the superparticle type, we will first
simplify the last four terms in (\ref{superpartaction}). For this, we note that
in ten spacetime dimensions the Dirac matrices are taken in a Majorana
representation, so that $\Gamma^0$ is antisymmetric while $\Gamma^i$,
$i=1,\dots,9$, are symmetric matrices~\cite{schwarz}. We also treat
$\vartheta,\dot\vartheta$ as an anticommuting pair of variables in the action
$\sf S$. Then, it is easy to check that the third term in
(\ref{superpartaction}) vanishes, because via an integration by parts it can be
written as
\beq
-\frac i{36}\,\int\limits_0^1d\tau~x^0\,\left(\dot{\vartheta}^\top\,\Gamma^0\,
u\slash\,\vartheta+\vartheta^\top\,\Gamma^0\,u\slash\,\dot\vartheta\right)
=0 \ ,
\label{3rdterm0}\eeq
where we have used the Dirac algebra to write
$\Gamma^0\,u\slash=-u\slash\,\Gamma^0$. In a similar way one readily checks
that the fourth and fifth terms in
(\ref{superpartaction}) are zero. By the same techniques one finds that the
last term is non-vanishing, and after some algebra it can be expressed in the
form $\frac
i4\,\int_0^1d\tau~\vartheta^\top\,x^j\,u^i\,\Gamma_{ij}\,\dot\vartheta$. The
action (\ref{superpartaction}) can therefore be written as
\beq
{\sf S}=\int\limits_0^1d\tau~\left[\,p_i\left(\dot x^i+i\,
\overline{\vartheta}\,\Gamma^i\,\dot\vartheta\right)-i\,\ell^\top\,\dot
\vartheta+\dots\right] \ ,
\label{superactionfinal}\eeq
where
\beq
p_i=u_i\,x^0 \ , ~~ \ell=x^i\,u^j\,\Gamma_{ij}\,\vartheta \ ,
\label{pietadef}\eeq
and we have rescaled the worldline spinor fields $\vartheta\mapsto2\vartheta$.

The action (\ref{superactionfinal}) is, modulo mass-shell constraints, that of
a twisted superparticle~\cite{hull}, which admits a manifestly covariant
quantization. The first term is the standard non-relativistic superparticle
action, while the inclusion of the fermionic field $\ell$ modifies the
canonically conjugate momentum to $\vartheta$ as
$\pi_\vartheta=p\slash\vartheta-\ell$. Note that the quantity $p_i$ in
(\ref{pietadef}) is the expected momentum of the uniformly moving D-particle,
while $\ell$ is proportional to its angular momentum. In the present case
$p_\mu\,p^\mu\neq0$, so that the supersymmetric impulse operator describes a
{\it massive}, non-relativistic twisted superparticle. The twist in fermionic
momentum
$\pi_\vartheta$ vanishes if there is no angular momentum, for instance if the
D-particle recoils in the direction of scattering. The equations of motion
which follow from the action (\ref{superactionfinal},\ref{pietadef}) can be
written as
\beq
\dot x^0=u_i\,\dot x^i=u\slash\,\dot\vartheta=0 \ ,
\label{finaleqsmotion}\eeq
which imply that $x^0$ and the components of $x^i$ and $\vartheta$ along the
direction of motion are independent of the proper time $\tau$. In general the
remaining components of $x^\mu$ and $\vartheta$ are $\tau$-dependent. These
classical configurations agree with the interpretation of the worldsheet zero
mode of the field $x^0$ as the target space time and also of the uniform motion
of the D-particles. In particular, the Galilean trajectory
$x^i(\tau)=y^i(\tau)+u^i\,x^0$, appropriate for the kinematics of a heavy
D0-brane, solves (\ref{finaleqsmotion}) provided that the component of the
vector $y^i(\tau)$ along the direction of recoil is independent of the
worldline coordinate $\tau$.

There are some important differences in the present case from the standard
superparticle kinematics. The action (\ref{superactionfinal}) generically
possesses a fermionic $\kappa$-symmetry defined by the transformations
\bea
\delta_\kappa\vartheta&=&p\slash\,\kappa \ , \nn\\\delta_\kappa
\ell&=&2\,p_i\,p^i\,\kappa \ , \nn\\\delta_\kappa x^i&=&
i\,\kappa\,p\slash\,\Gamma^i\,\vartheta \ ,
\label{kappasym}\eea
where $\kappa$ is an infinitesimal Grassmann spinor parameter. It is also
generically invariant under a twisted ${\cal N}=2$ super-Poincar\'e
symmetry~\cite{hull}. However, the choices (\ref{pietadef}) break these
supersymmetries, which is expected because the D-brane motion induces a
non-trivial vacuum energy. The configurations (\ref{pietadef}) of course arise
from the geodesic bosonic paths in the non-relativistic limit $u_i\ll1$, or
equivalently in the limit of heavy BPS mass for the D-particles, which is the
appropriate limit to describe the tree-level dynamics here. The Galilean
solutions of (\ref{finaleqsmotion})
described above explicitly break the $\kappa$-symmetry (\ref{kappasym}).

Therefore, we see that the supersymmetric completion
of the impulse operator (for weakly-coupled strings) describes the dynamics of
a twisted supersymmetric D-particle in the non-relativistic limit, with a
gauge-fixing that breaks its target space supersymmetries. In turn, this broken
supersymmetry implies that the vertex operator (\ref{vertexDsusy}) does not
generate the action of the super-Poincar\'e group on the brane, and
consequently the super-D-particle does not evolve in target space according to
super-Galilean
transformations~\cite{galilei}. The structure of the worldsheet logarithmic
superconformal algebra is such that these spacetime properties of D-brane
dynamics are enforced by the impulse operators.

\newsection{Modular Behaviour}

In the case of string solitons, the non-trivial mixing between the logarithmic
$C_\epsilon$ and $D_\epsilon$ operators leads to logarithmic modular
divergences in bosonic annulus amplitudes, and it is associated with the lack
of unitarity of the low-energy effective theory in which quantum D-brane
excitations are neglected~\cite{KM,kmw,MS1}. We now examine how these features
are modified in the presence of the logarithmic ${\cal N}=1$ superconformal
pair. For this, we consider the open superstring propagator between two
scattering states $|{\cal E}_\alpha\rangle$ and $|{\cal E}_\beta\rangle$,
\beq
\triangle_{\alpha\beta}=\langle{\cal E}_\alpha|\frac1{L_0-1/2}
|{\cal E}_\beta\rangle=-\int\limits_{\cal F}
\frac{dq}q~\langle{\cal E}_\alpha|q^{L_0-1/2}|{\cal E}_\beta\rangle \ ,
\label{stringpropgen}\eeq
where the Virasoro operator $L_0$ is defined through the Laurent expansion of
the energy-momentum tensor $T(z)=\sum_nL_n\,z^{-n-2}$, and the factor of
$\frac12$ is the normal ordering intercept in the Neveu-Schwarz sector. Here
$q=\e^{2\pi i\tau}$, with $\tau$ the modular parameter of the worldsheet strip
separating the two states $|{\cal E}_\alpha\rangle$ and $|{\cal
E}_\beta\rangle$, and $\cal F$ is a fundamental modular domain of the complex
plane. We shall
ignore the superconformal ghosts, whose contributions would not affect the
qualitative results which follow.

For the purely bosonic string, divergent contributions to the modular integral
would come from a discrete subspace of string states of vanishing conformal
dimension corresponding to the spectrum of linearized fluctuations in the
soliton background~\cite{FPR1,KM,MS1}. Since in the
present case these are precisely the states associated with the logarithmic
recoil operators, we should analyse carefully their contributions to the
propagators (\ref{stringpropgen}). We introduce the highest weight states
$|\phi\rangle=\phi(0)|0\rangle$,
$\phi=C_\epsilon,D_\epsilon,\chi^{~}_{C_\epsilon},\chi^{~}_{D_\epsilon}$, with
the understanding that the $\partial^{~}_{\!\perp} x^i$ and $\psi^i$ parts of
the vertex operator (\ref{vertexDsusy}) are included. This has the overall
effect of replacing $\Delta_\epsilon$ in the bosonic parts of the operator
product expansions everywhere by the anomalous dimension
$h_\epsilon=1+\Delta_\epsilon$ of the impulse operator, while in the fermionic
parts $\Delta_\epsilon+\frac12$ is replaced everywhere by $h_\epsilon$. Using
(\ref{TCD}) and (\ref{SUSYOPE}), the $2\times2$ Jordan cell decompositions of
the bosonic and fermionic Virasoro generators are then given by
\beq
\new{\begin{array}{rrlrrl}
L^{\rm b}_0|C_\epsilon\rangle&=&h_\epsilon|C_\epsilon\rangle \ , ~~&
L^{\rm b}_0|D_\epsilon\rangle&=&h_\epsilon|D_\epsilon\rangle+|C_\epsilon
\rangle \ , \\L^{\rm f}_0|\chi^{~}_{C_\epsilon}\rangle&=&h_\epsilon|
\chi^{~}_{C_\epsilon}\rangle \ , ~~&L^{\rm f}_0|\chi^{~}_{D_\epsilon}
\rangle&=&h_\epsilon|\chi^{~}_{D_\epsilon}\rangle+|\chi^{~}_{C_\epsilon}
\rangle \ , \end{array}}
\label{L0Jordan}\eeq
where $L_0=L_0^{\rm b}+L_0^{\rm f}$. Using the factorization of bosonic and
fermionic states, in the Jordan blocks spanned by the logarithmic operators we
have~\cite{flohr3}
\beq
q^{L_0}\,|C_\epsilon,D_\epsilon\rangle\otimes|\chi^{~}_{C_\epsilon},
\chi^{~}_{D_\epsilon}\rangle=q^{h_\epsilon}\pmatrix{1&0\cr\ln q&1\cr}
|C_\epsilon,D_\epsilon\rangle\otimes q^{h_\epsilon}\pmatrix{1&0\cr\ln q&1\cr}
|\chi^{~}_{C_\epsilon},\chi^{~}_{D_\epsilon}\rangle \ .
\label{qL0Jordan}\eeq
The corresponding expectation value (\ref{stringpropgen}) in such a state is
then given by
\beq
\triangle_{CD}=-\int\limits_{\cal F}dq~q^{2\Delta_\epsilon+1/2}\,\langle
C_\epsilon,D_\epsilon|\pmatrix{1&0\cr\ln q&1\cr}|C_\epsilon,D_\epsilon\rangle\,
\langle\chi^{~}_{C_\epsilon},\chi^{~}_{D_\epsilon}|\pmatrix{1&0\cr\ln q&1\cr}
|\chi^{~}_{C_\epsilon},\chi^{~}_{D_\epsilon}\rangle \ .
\label{propCD}\eeq
The dangerous region of moduli space is ${\rm Im}\,\tau\to+\infty$, in which
$q\sim\delta\to0^+$. Using $\Delta_\epsilon=0$ as $\epsilon\to0^+$, we can
easily check that the contributions to the modular integration in
(\ref{propCD}) from this region {\it vanish}. For instance, the worst behaviour
comes from the term in the integrand involving $\sqrt q\,(\ln q)^2$, which upon
integration over a small strip ${\cal F}_\delta$ of width $\delta$ produces a
factor
\beq
\int\limits_{{\cal F}_\delta}
dq~\sqrt q\,(\ln q)^2\simeq\frac23\,\delta^{3/2}\,\left(
(\ln\delta)^2-\frac43\,\ln\delta+\frac89\right) \ ,
\label{stripint}\eeq
which vanishes in the limit $\delta\to0^+$. Therefore, in quantities involving
matrix elements of the string propagator in logarithmic states, the
incorporation of worldsheet superconformal partners cancels the modular
divergences that are present in the purely bosonic case. It is also
straightforward to arrive at this conclusion in the Ramond sector of the
superstring theory. Notice that although the explicit calculation above is
carried out with respect to the chosen basis (\ref{L0Jordan}) within the Jordan
cell, the same qualitative conclusion is arrived at under any change of basis
$|C_\epsilon,D_\epsilon\rangle\to|aC_\epsilon+bD_\epsilon,cC_\epsilon+d
D_\epsilon\rangle$. This is because the strip integral (\ref{stripint}) is the
worst behaved one and any change of basis will simply mix it with better
behaved modular integrals. Furthermore, physical string scattering amplitudes
will involve the superstring propagator with sums over complete sets in an
invariant, basis-independent form. Its effect on such physical quantities is
therefore independent of the chosen base.

This cancellation of infinities has dramatic consequences for the behaviours of
higher genus amplitudes. In the purely bosonic case, where the modular
divergences persist, the logarithmic states yield non-trivial contributions to
the sum over string states and imply that, to leading order, the genus
expansion is dominated by contributions from degenerate Riemann surfaces whose
strip sizes become infinitely thin~\cite{FPR1,KM,MS1}. Such amplitudes can be
described in terms of bi-local worldsheet operators and the truncated
topological series can be summed to produce a non-trivial probability
distribution on the moduli space of running coupling constants of the slightly
marginal $\sigma$-model~\cite{MS1}. The functional Gaussian distribution has
width proportional to $\sqrt{\ln\delta}$, and the string loop divergences are
cancelled by a version of the Fischler-Susskind mechanism. However, we see here
that this structure
disappears completely when one considers the full superstring theory. This
means that in the supersymmetric case one has to contend with the full genus
expansion of string theory which is not even a Borel summable series. The
dominance of pinched annular surfaces, as well as the loss of unitarity due to
the logarithmic mixing, can now be understood as merely an artifact of the
tachyonic instability of the bosonic string. Once the appropriate
superconformal partners to the logarithmic operators are incorporated, the
theory is free from divergences, at least at the level of string loop
amplitudes. Heuristically, this feature can be understood from the form of the
fermionic two-point functions (\ref{SUSY2pt}), which for $\Delta=0$ reduce to
conventional fermionic correlators with no logarithmic scaling violations on
the worldsheet. The zero dimension fermion fields, after incorporating the
worldsheet superconformal ghost fields, thereby have the usual effect of
removing instabilities from the theory.

\newsection{The Zamolodchikov Metric}

Another way to understand the effect of the fermionic fields in the recoil
problem is through the Zamolodchikov metric in the sector corresponding to the
logarithmic states. It is defined by the short-distance two-point functions
\beq
{\cal G}_{\phi\phi'}=\Lambda^{2h}\,\lim_{z\to w}\,\Bigl\langle\phi(z)\,
\phi'(w)\Bigr\rangle \ , ~~ \phi,\phi'=C,D,\chi^{~}_C,\chi^{~}_D \ ,
\label{Zamdef}\eeq
and by using (\ref{CD2pt}) and (\ref{SUSY2pt}) it can be represented as the
$4\times4$ matrix
\beq
{\cal G}=\pmatrix{0&\xi&0&0\cr\xi&d-2\xi\ln\Lambda&0&0\cr
0&0&0&2\Delta\xi\cr0&0&2\Delta\xi&2(\xi+\Delta d-2\Delta\xi
\ln\Lambda)\cr} \ .
\label{Zam4by4}\eeq
In the upper left bosonic $2\times2$ block we find a logarithmically divergent
term, which may be associated to the logarithmic modular divergences that are
present in the bosonic case. On the other hand, in the lower right fermionic
$2\times2$ block we find that the logarithmic divergence generically appears
only through the term which is proportional to $\Delta\ln\Lambda$. For the
recoil problem, in which the conformal dimension of the operators is correlated
with the worldsheet ultraviolet scale through the relations (\ref{Deltavarep})
and (\ref{epsilonLambdarel}), this term is a finite constant. Thus, in contrast
to its bosonic part, the fermionic part of the Zamolodchikov metric is
scale-invariant. This is just another reflection of the fact that the fermionic
logarithmic operators do not themselves lead to any logarithmic divergences and
act to cure the bosonic string theory of its instabilities. In fact, this
property on its own is motivation for the identification
(\ref{epsilonLambdarel}) of worldsheet and target space regularization
parameters which was used to arrive at the logarithmic conformal algebra. In
turn, this correlation is then also consistent with the Galilean non-invariance
(\ref{chiCDscale}) which derives from the twisted superparticle interpretation
of the previous section. Nevertheless, the vanishing correlation functions in
(\ref{CD2pt}) and (\ref{SUSY2pt}) indicate the existence of a hidden
supersymmetry in the dynamics of moving D-branes. For instance, it is
straightforward to check that the fermionic Noether supercurrents associated
with spatial translations induce the same logarithmic scaling violations that
the bosonic ones do~\cite{FPR1,KM}.

It is curious to note that the Zamolodchikov metric (\ref{Zam4by4}) becomes
degenerate in the conformal limit $\Delta\to0$, which corresponds to the
infrared fixed point of the worldsheet field theory (in the sense that the size
of the worldsheet is infinite in units of the ultraviolet cutoff). In this
limit all two-point correlation functions involving the fermionic field
$\chi^{~}_C$ vanish. Whether or not this implies that $\chi^{~}_C$ completely
decouples from the theory requires knowledge of higher order correlators of
the theory. Furthermore, in that case there are no logarithmic scaling
violations, since $\langle\chi^{~}_D(z)\,\chi^{~}_D(w)\rangle=2\xi/(z-w)$ in
the limit $\Delta\to0$. Generally, the vanishing of two-point functions in a
logarithmic
conformal field theory implies some special properties of the model. In the
purely bosonic cases, it is known that such a vanishing property is associated
with the existence of hidden symmetries corresponding to some conserved
current~\cite{CKT}. A similar situation may occur in the supersymmetric case,
indicating the presence of some new fermionic symmetries. For this to be case,
there must be some other field to which the field $\chi^{~}_C$ couples. While
the extra hidden symmetry may be related to the fact that $\chi_C^{~}$ is a
null field in the subspace of primary fields, it should not be a true null
field. This interesting issue deserves further investigation. Notice however
that in the recoil problem, the pertinent correlation functions are
non-vanishing in the slightly-marginal case where $\epsilon\neq0$.

Notice also that the
degeneracy of the metric (\ref{Zam4by4}) in the limit $\Delta\to0$ may not be a
true singularity of the moduli space of coupling constants. To further
elaborate on this point requires computation of the corresponding curvature
tensor, and its associated invariants which, being invariant under changes of
renormalization group scheme, contain the true physical information of the
theory. However, this again requires knowledge of the three-point and
four-point correlation functions among the pertinent vertex operators, which at
present
are not available.

The Zamolodchikov metric is also a very important ingredient in the
construction of the effective target space action of the theory. In the bosonic
case such a moduli space action reproduces the Born-Infeld action for the
D-brane dynamics in the Neumann representation~\cite{MS1}. The
supersymmetrization of the worldsheet theory along the lines of section~4 will
produce the same
Born-Infeld action, with the only effect that the tachyonic instabilities are
again removed and no renormalization of the coupling constants are
required~\cite{Tseyt1}. This is immediate due to the form of (\ref{Zam4by4}).
On the other hand, the target space supersymmetrization of the Born-Infeld
action in ten
dimensions is known~\cite{Tseyt1}. The photino field $\lambda$ corresponds to
the Goldstino particle of the super-Poincar\'e symmetry which is spontaneously
broken by the presence of the D-brane. The resulting action does however
possess local spacetime $\kappa$-symmetry. We may then expect an appropriate
version of this action to emerge within the target space formalism of the
previous section, with corresponding breaking of the fermionic
$\kappa$-symmetry.

\subsection*{Acknowledgments}

R.J.S. thanks J.~Figueroa-O'Farrill for helpful discussions. The work of R.J.S.
was supported in part by an Advanced Fellowship from the Particle Physics and
Astronomy Research Council~(U.K.).


\begin{thebibliography}{99}

\baselineskip=12pt

\bibitem{FPR1} W.~Fischler, S.~Paban and M.~Rozali, Phys. Lett. {\bf B352}
(1995) 298 [{\tt hep-th/9503072}].

\bibitem{KM} I.I.~Kogan and N.E.~Mavromatos, Phys. Lett. {\bf B375} (1996) 111
[{\tt hep-th/9512210}].

\bibitem{Gurarie} V.~Gurarie, Nucl. Phys. {\bf B410} (1993) 535 [{\tt
hep-th/9303160}].

\bibitem{CKT} J.-S.~Caux, I.I.~Kogan and A.M.~Tsvelik, Nucl. Phys. {\bf B466}
(1996) 444 [{\tt hep-th/9511134}].

\bibitem{recoil} V.~Periwal and \O.~Tafjord, Phys. Rev. {\bf D54} (1996) 3690
[{\tt hep-th/9603156}]; D.~Berenstein, R.~Corrado, W.~Fischler, S.~Paban and
M.~Rozali, Phys. Lett. {\bf B384} (1996) 93 [{\tt hep-th/9605168}]; J.~Ellis,
N.E.~Mavromatos and D.V.~Nanopoulos, Int. J. Mod. Phys. {\bf A12} (1997) 2639
[{\tt hep-th/9605046}].

\bibitem{kmw} I.I.~Kogan, N.E.~Mavromatos and J.F.~Wheater, Phys.\ Lett.\ {\bf
B387} (1996) 483 [{\tt hep-th/9606102}].

\bibitem{MS1} N.E.~Mavromatos and R.J.~Szabo, Phys. Rev. {\bf D59} (1999)
104018 [{\tt hep-th/9808124}].

\bibitem{cardy} J.L.~Cardy, Nucl.\ Phys.\ {\bf B240} [FS12] (1984) 514; {\bf
B324} (1989) 581; J.L.~Cardy and D.C.~Lewellen, Phys.\ Lett. {\bf B259} (1991)
274.

\bibitem{CKLT} J.-S.~Caux, I.I. Kogan, A. Lewis and A.M. Tsvelik, Nucl. Phys.
{\bf B489} (1997) 469 [{\tt hep-th/9606138}].

\bibitem{susylog} M.~Khorrami, A.~Aghamohammadi and A.M.~Ghezelbash, Phys.\
Lett.\ {\bf B439} (1998) 283 [{\tt hep-th/9803071}].

\bibitem{susylogfrac} F.~Kheirandish and M.~Khorrami, Eur. Phys. J. {\bf C18}
(2001) 795 [{\tt hep-th/0007013}]; Eur. Phys. J. {\bf C20} (2001) 593 [{\tt
hep-th/0007073}].

\bibitem{hull} W. Siegel, Class. Quant. Grav. {\bf 2} (1985) L95;
Nucl. Phys. {\bf B263} (1985) 93; M.B.~Green and C.M.~Hull, Nucl. Phys. {\bf
B344} (1990) 115; C.M.~Hull and J.-L.~V\'azquez-Bello,
Nucl.\ Phys.\ {\bf B416} (1994) 173 [{\tt hep-th/9308022}].

\bibitem{RTAK} M.R.~Rahimi~Tabar, A.~Aghamohammadi and M.~Khorrami, Nucl. Phys.
{\bf B497} (1997) 555 [{\tt hep-th/9610168}]; M.A.I.~Flohr, Nucl. Phys. {\bf
B514} (1998) 523 [{\tt hep-th/9707090}].

\bibitem{flohr1} M.A.I.~Flohr, [{\tt hep-th/0009137}].

\bibitem{GabKausch} M.R.~Gaberdiel and H.G.~Kausch, Phys. Lett. {\bf B386}
(1996) 131 [{\tt hep-th/9606050}]; Nucl. Phys. {\bf B538} (1999) 631 [{\tt
hep-th/9807091}].

\bibitem{flohr2} M.A.I.~Flohr, [{\tt hep-th/0107242}].

\bibitem{Dmoving} J.~Dai, R.G.~Leigh and J.~Polchinski, Mod. Phys. Lett. {\bf
A4} (1989) 2073; R.G.~Leigh, Mod. Phys. Lett. {\bf A4} (1989) 2767; C.G.~Callan
and I.R.~Klebanov, Nucl. Phys. {\bf B465} (1996) 473 [{\tt hep-th/9511173}];
M.~Li, Nucl. Phys. {\bf B460} (1996) 351 [{\tt hep-th/9512042}].

\bibitem{kw} S.~Moghimi-Araghi and S.~Rouhani, Lett. Math. Phys. {\bf 53}
(2000) 49 [{\tt hep-th/0002142}]; I.I.~Kogan and J.F.~Wheater, Phys.\ Lett.\
{\bf B486} (2000) 353 [{\tt hep-th/0003184}]; Y.~Ishimoto, [{\tt
hep-th/0103064}]; S.~Kawai and J.F.~Wheater, Phys.\ Lett.\ {\bf B508} (2001)
203 [{\tt hep-th/0103197}].

\bibitem{GSW} M.B. Green, J.H. Schwarz and E. Witten, {\it Superstring Theory}
(Cambridge University Press, 1987).

\bibitem{Bachas} C.P.~Bachas, Phys. Lett. {\bf B374} (1996) 37 [{\tt
hep-th/9511043}].

\bibitem{schwarz} J.H. Schwarz, in: {\it Superstrings and Supergravity}, eds.
A.T. Davies and D.G. Sutherland (SUSSP Publications, 1986).

\bibitem{galilei} R. Puzalowski, Acta Phys. Austriaca {\bf 50} (1978) 48.

\bibitem{flohr3} M.A.I.~Flohr, Int. J. Mod. Phys. {\bf A11} (1996) 4147 [{\tt
hep-th/9509166}].

\bibitem{Tseyt1} A.A. Tseytlin, in: {\it The Many Faces of the Superworld}, ed.
M.A. Shifman (World Scientific, 2000), p. 417 [{\tt hep-th/9908105}].

\end{thebibliography}
\end{document}